\documentstyle[preprint,epsfig,aps]{revtex}
\title{Complexation of DNA with Cationic Surfactant }
\author{ Paulo S. Kuhn, Marcia C. Barbosa, 
and Yan Levin\footnote{Corresponding author;
e-mail: levin@if.ufrgs.br}\\
Instituto de F{\'\i}sica, Universidade Federal
do Rio Grande do Sul\\ Caixa Postal 15051, CEP 91501-970, Porto Alegre, RS, Brazil\\ }
\begin{document}
\maketitle
\begin{abstract}

Transfection of an anionic polynucleotide through a negatively
charged 
membrane is an important problem in genetic engineering. The direct
association of cationic surfactant to DNA
decreases the effective negative charge of the nucleic  acid,
allowing  
the DNA-surfactant complex to approach a negatively charged membrane.
The paper develops a theory for solutions composed of 
polyelectrolyte, salt, and
ionic surfactant. The theoretical predictions are compared 
with the experimental measurements.

\end{abstract}

\bigskip

PACS.05.70.Ce - Thermodynamic functions and equations of state

PACS.61.20.Qg - Structure of associated liquids: electrolytes, molten 
salts, etc.

PACS.61.25.Hq - Macromolecular and polymer solutions; polymer melts; swelling

\newpage

Gene therapy has increasingly
captured public attention 
after the first gene transfer study in humans 
was completed in 1995\cite{Blae,Hope,And}.
The  procedure  delivers a functional
polynucleotide sequence into the cells of an organism affected by
a genetic disorder. The gene delivery system that has been adopted in
over $ 90\%$
of the clinical trials to date is in the form of genetically engineered
non-replicating retroviral or  adenoviral
vectors\cite{Hope}. Unfortunately,
the adverse response
of the immune system has hindered application of virus based gene therapy. 
New strategies are now being explored
\cite{Felg2,FelgRin,Fried,Felg,Lan1,Ver,Croo,Hodg,Akh,Ben,Sul,Jul,Gao,Lew,Far,Lasic}. One of the approaches,  
pioneered by Felgner and Ringold\cite{FelgRin}, relies on association between the
anionic nucleic acid and cationic lipid liposomes.  
The process of association neutralizes the excess negative charge of
a polynucleotide, allowing the DNA-lipid complex
to approach a negatively charged phospholipid membrane.  
Unfortunately, the  cationic lipids 
and surfactants are toxic to an organism. 
A  question that we will try to answer in this
letter is: What is the minimum
amount of  cationic surfactant or lipid that 
is necessary to form a complex and how does this amount depends 
on various properties of a  system?

We study a solution consisting  of  DNA
segments of density $\rho_{\rm DNA}$,  
surfactants of density $\rho_{\rm surf}$, and salt  molecules of density
 $\rho_{\rm salt}$\cite{KLB}. The solvent  is idealized as a uniform
 medium of dielectric constant $D$.
 Since the DNA molecule has a large intrinsic rigidity,
we   model it as a cylinder of 
fixed length and diameter.
When in solution, the $Z$ phosphate groups of the DNA strand become
ionized, resulting in a net molecular charge, $-Zq$. An equivalent
number of counterions of density $Z\rho_{\rm DNA}$ are released
into solution preserving the overall charge neutrality.
Similarly, the cationic surfactant molecule in aqueous solution becomes
ionized, producing a free 
negative ion and a flexible
chain consisting of one
positively charged hydrophilic head group and 
a neutral hydrophobic tail.
The ions of salt, the 
counterions, and the negative ions dissociated from the surfactant
are modeled as hard spheres with point charge located
at the center. For simplicity, we shall call the  negative
ions,  ``coions'',  and the  positive ions, ``counterions'' ---  independent
 of the species from which they are derived 
(see Figure $1$).  

The strong 
electrostatic attraction between the  
 counterions, cationic surfactant, and the
 DNA favors  formation of clusters consisting of one  DNA molecule
and  $n_{\rm count}$ associated counterions, and 
$n_{\rm surf}$ associated surfactants.
The  process of association
neutralizes  $n_{\rm surf}+n_{\rm count}$ phosphate groups
of a  DNA molecule, decreasing the
net charge of a complex to,
$q_{\rm complex}=-(Z-n_{\rm surf}-n_{\rm count})q$ (Figure $1$). 
Our task is to determine the values of  $n_{\rm count}$ and  
$n_{\rm surf}$ which
are thermodynamically favored, i.e. 
which minimize the overall Helmholtz
free energy of solution.

For a dilute suspension, the main contributions
to the free energy  can be subdivided into three parts: the  energy
that it takes to construct an {\it isolated} complex, $F_{\rm association}$;
 an energy
that it takes to solvate this complex in the ionic sea, 
$F_{\rm solvation}$;  and  the entropic energy of  
mixing, $F_{\rm mixing}$.

To calculate the free energy of an isolated cluster, we
use the following simplified  model of a complex. 
Each monomer of a polyion is treated as free or occupied by 
a counterion or a surfactant (Figure $2$). 
We associate with each monomer $i$
two occupation variables $\sigma(i)$ and 
$\tau(i)$, such
that $\sigma(t)=1$ if the site is occupied by a
condensed  counterion, and $\sigma(i)=0$  otherwise.  The 
occupation
number for surfactants, $\tau(i)$, behaves in a similar way.
The free energy can now be calculated as a logarithm
of the  Boltzmann sum over all possible configurations of 
condensed counterions
and surfactants along the polyion,
%----------------------------------------
\begin{eqnarray}
\beta F_{\rm association}&=&
- \ln \sum_{\nu}^{} e^{-\beta E_{\nu} }  \;.
\end{eqnarray}
%-----------------------------------------
The energy of a given configuration $\nu$ can be subdivided into two parts,
$E_{\nu}=E_{1}+E_{2}$, where the electrostatic contribution is,
\begin{equation}
E_{1}= \frac {q^2} {2} \sum_{i\neq j} \frac {[-1+\sigma(i)+\tau(i)] 
[-1+\sigma(j)+\tau(j)]} {D |r(i)-r(j)|}\; .
\end{equation} 
%-----------------------------------------
The energy $E_2$ arises from hydrophobicity of surfactant molecules.
Clearly, when two adjacent sites are occupied by surfactants,
 the net exposure of hydrocarbon tails to water is
reduced. We capture this effect by introducing an additional
contribution to the overall energy of interaction, $E_2$, given by
%--------------------------------------
\begin{equation}
E_{2}=- \frac{\chi}{2} \sum_{< i\neq j >}\tau(i)\tau(j)\; ,
\end{equation}
%----------------------------------------
where the sum runs over the nearest neighbors. The parameter $\chi$, related 
to the decrease in overall energy 
due to the agglomeration of surfactants, is obtained  
from an independent experimental
measurement of 
the energy 
that is required to move
one surfactant molecule
from a monolayer to bulk\cite{Isra}.

The exact solution of even  
this one-dimensional ``sub-problem''
is rather difficult due to the 
long-ranged character of  
the Coulomb force. To proceed we could use a
 mean-field 
approximation, but while
the mean-field theory works very well for long-range potentials,
for one-dimensional
systems with short-range forces, it can lead to unphysical
instabilities \cite{KLB}. In order to avoid this difficulty, we
treat the long-range electrostatic part of the association free energy 
using a mean-field approximation,
while performing an exact calculation for the 
short-range hydrophobic interaction.

Once a cluster, constructed in isolation, is
introduced  into solution, it gains
an additional energy
due to electrostatic interactions with the other entities.
The free energy gained in the process of solvation
can be obtained using the Debye-H{\"u}ckel \cite{KLB,DH}
theory.  Let us  fix the position of one cluster
 and ask  what is the electrostatic
potential $\Phi$  that this cluster 
feels as a result of the presence of all the other
clusters, surfactants, counterions, and coions. To answer  
this question, consider the Poisson equation,
%----------------------------------------
\begin{eqnarray}
\nabla^2 \Phi &=& - \frac{4 \pi}{D} \rho_q \;.
\end{eqnarray}
%----------------------------------------
To make the problem well posed, this equation must be provided
with a closure which would relate the electrostatic potential,
$\Phi$,
to the net charge density, $\rho_q$. A simple closure motivated
by ideas derived from the  Debye-H{\"u}ckel theory is to suppose that  
the free (unassociated) surfactants, counterions, and  coions are distributed
around the complex in accordance with the Boltzmann distribution, 
with other clusters providing a neutralizing background, 
%----------------------------------------
\begin{eqnarray}
\rho_q & = & q_{\rm complex} \rho_{\rm DNA} 
+  q \rho_{\rm count} e^{-\beta q \Phi} 
- q \rho_{\rm coion} e^{ +\beta q \Phi } 
+ q \rho_{\rm surf} e^{-\beta q \Phi} ,
\end{eqnarray} 
%-------------------------------------------
where $\beta=1/(k_B T)$.

Inserting (5) into (4) we obtain
the Poisson-Boltzmann equation, which after linerization
reduces to the familiar Helmholtz form.
The linearization is justified since all the non-linearities are
effectively included in the renormalization of DNA charge
by the  
formation of clusters\cite{Lev,Alex,FL}.  The Helmholtz equation
can be solved analytically, yielding the electrostatic
 potential of a complex.  
The electrostatic free energy of solvation is obtained
through the usual Debye charging process\cite{DH,KLB}.
 
The free energy due to mixing of various species
is a sum of individual
entropic contributions,
%---------------------------------------------
\begin{equation}
F_{mixing}=F_{counterion}+F_{coion}+F_{surfactant}+F_{complex}.
\end{equation} 
%----------------------------------------------
The structure of each one
of these terms is similar to that of an ideal gas and can be
calculated using the Flory's theory
of polymer melts\cite{Flo}. For
example,  in the case of  coions, the reduced free energy density is
$\beta F_{\rm coion}/V=\rho_{\rm coion}\ln (\phi_{\rm coion}/\zeta)-\rho_{\rm coion}$,
where $\phi_{\rm coion}$  is the volume
fraction occupied by the negative microions,
 while $\zeta$ is a factor that takes
into account the internal structure of each specie. For
structureless particles such as coions and counterions, $\zeta=1$.
For a flexible linear surfactant chain, $\zeta$ is the number of
monomers comprising a molecule \cite{Flo}.
For a  complex  made of a  
DNA segment and condensed surfactants and counterions,
$\zeta$ is related to the number of different configurations
which can arise when 
$n_{count}$ counterions
and $n_{surf}$ surfactant molecules 
associate to a DNA molecule. 
Minimizing the total free energy,
%-----------------------------------
\begin{equation}
F=F_{\rm mixing}+F_{\rm association}+F_{\rm solvation}  ,
\end{equation}
%-----------------------------------------------
with respect to the number of associated  counterions and
 surfactants,
%----------------------------------
\begin{equation} 
\frac{\partial F}{\partial n_{\rm cout}}
=\frac{\partial F}{\partial n_{\rm surf}}=0,
\end{equation}
%--------------------------------------
we find the thermodynamically
preferred values for the number of condensed 
particles.
We define a ``surfoplex''  to be  a complex in which almost 
all of the DNA's phosphate groups are neutralized by
the associated surfactant molecules.  As mentioned in the
introduction, we are interested in the minimum amount of cationic
surfactant needed to transform  naked DNA into surfoplexes.  To this effect,
we study the dependence of the number of condensed surfactant
molecules, $n_{\rm surf}$, on the bulk concentration of surfactant.  
Figure 3 demonstrates
the location of the cooperative
binding transition associated with the  formation of surfoplexes.  
The transition is a result
of competition between the entropic,
the electrostatic, and the hydrophobic
interactions.  Clearly, for small concentrations of surfactant,
binding to the polyions is not thermodynamically favored, since the
system can lower its free energy by keeping the surfactant in
the bulk, thus gaining entropy. As the density of surfactant increases,
the gain in electrostatic energy, due to binding, outweighs the loss
of entropy due to confinement which, in any case, is largely compensated 
by the release of bound counterions\cite{Man1,Man2}. The cooperativety 
predicted by our theory and observed in experiments\cite{Daw,Shi} is due to
the fact that once the first surfactant is bound, the binding of
additional surfactants is strongly favored by the decrease in hydrophobic
energy of the exposed hydrocarbon tails. The high degree of cooperativety
allows us to clearly define how much surfactant is necessary to 
form a surfoplex.
To check the predictions of our theory, we compare it to
the recent experimental measurements\cite{Daw}  of 
binding isotherm in a solution of DNA, 
dodecyltrimethylammonium bromide, 
and  salt, (Figure $3$). 
The agreement is  encouraging specially since the
 theory {\it  does not have any fitting parameters}.

{\bf ACKNOWLEDGMENTS}

\bigskip

We  would like to acknowledge helpful conversations 
with Profs. K.A. Dawson, Michael E. Fisher, 
and Prof. H. E. Stanley. This work was 
supported in part by CNPq - Conselho Nacional de
Desenvolvimento Cient{\'\i}fico e Tecnol{\'o}gico and FINEP -
Financiadora de Estudos e Projetos, Brazil.

%\newpage

\newpage

\begin{figure}[h]
\centerline{\epsfig{file=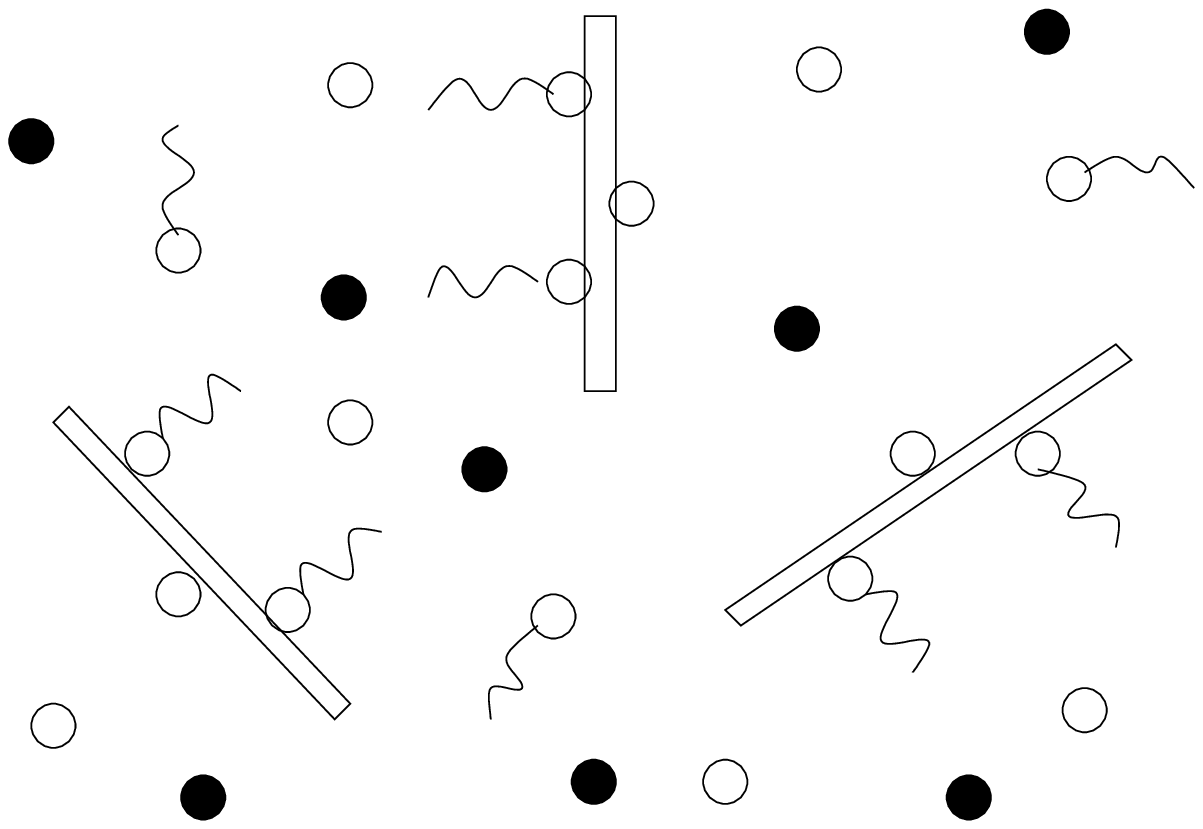,width=5cm,angle=270} }
\caption{{\it Schematic Illustration of the
Model}: Complexes formed by DNA molecules (rectangles), counterions
(solid circles) and flexible surfactant molecules ( open circles
with chains) are surrounded by free counterions, 
coions ( open circles), and  surfactant molecules.}
%\label{...}
\end{figure}
\begin{figure}[h]
\centerline{\epsfig{file=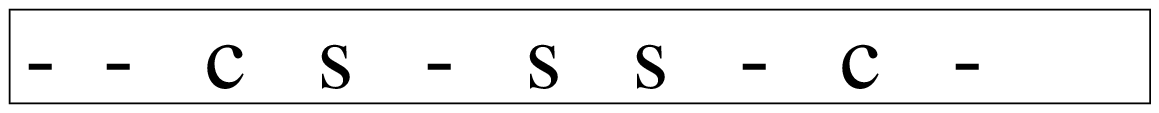,width=1cm,height=5cm,angle=270} }
\caption{{\it  Lattice model for forming a cluster.}
Each monomer can be either occupied by a counterion $(C)$ , a
surfactant $(S)$,
or neither $(-)$. Interaction between the counterions 
is purely electrostatic, while two neighboring surfactant
molecules gain energy due to hydrophobic effects.  }
%\label{...}
\end{figure}

\begin{figure}[h]
\centerline{\epsfig{file=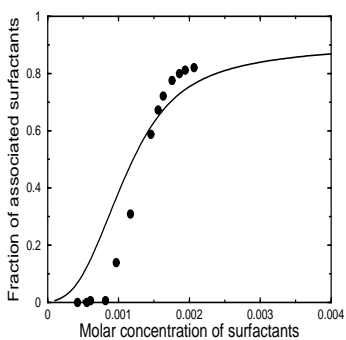,width=5cm,height=5cm} }
\caption{{\it  Fraction of monomers along the  DNA segment
associated with  surfactant molecules,  $n_{\rm surf}/Z$,
as a function of bulk density of surfactant, $\rho_{\rm s}$}.
The solid line is derived from the solution
of Eq.(8), while the
circles are the experimental data[29].
The experimental value of the hydrophobicity parameter is 
$\chi=-3.5k_BT$[20]. The concentrations of DNA and 
of added salt are $2 \times 10^{-6}M$ 
and $18 mM$, respectively. The length of a  DNA segment
 is $220$ base pairs.}
%\label{...}
\end{figure}
\end{document}